# Biomolecular imaging and electronic damage using X-ray free-electron lasers

**Harry M Quiney and Keith A Nugent**

ARC Centre of Excellence for Coherent X-ray Science, School of Physics, The University of Melbourne, Vic., 3010 AUSTRALIA

**Proposals to determine biomolecular structures from diffraction experiments using femtosecond X-ray free-electron laser (XFEL) pulses involve a conflict between the incident brightness required to achieve diffraction-limited atomic resolution and the electronic and structural damage induced by the illumination. Here we show that previous estimates of the conditions under which biomolecular structures may be obtained in this manner are unduly restrictive, because they are based on a coherent diffraction model that is not appropriate to the proposed interaction conditions. A more detailed imaging model derived from optical coherence theory and quantum electrodynamics is shown to be far more tolerant of electronic damage. The nuclear density is employed as the principal descriptor of molecular structure. The foundations of the approach may also be used to characterize electrodynamical processes by performing scattering experiments on complex molecules of known structure.**

The challenges involved in determining the structures of molecules to atomic resolution in non-crystalline samples using X-ray free-electron laser pulses are formidable[1]. Significant advances have already been made, however, in the design and preparation of experiments using fourth-generation sources[2] and the corresponding structural analysis of diffraction data. The difficulties posed by diffraction data collected from single molecules dropped with random orientations into the path of a freely propagating XFEL pulse have led to the development of an ingenious Bayesian characterization scheme by which two-dimensional diffraction patterns may be assembled into complete three-dimensional diffraction sets[3]. Such approaches enable the accumulation of diffraction data corresponding to identical targets, increasing the signal-to-noise ratio, particularly in the large-angle scattering sector of the data that carries the high-resolution information about molecular structure. The inevitable Coulomb explosion of molecules subjected to such pulses has also been addressed by detailed computational modeling of the interaction dynamics[1, 4, 5]. This has led to proposals for the incorporation of sacrificial tampers to delay the Coulomb explosion of the scattering target[6] or temporal gating of data acquisition to reduce the degradation of the structural information due to the molecular disintegration [7].

The feasibility of successful molecular structure determination from diffraction data using XFEL sources has been assessed in each of these studies by calculating some variant of the crystallographic $R$-factor,

$$R = \frac{\sum_{\mathbf{u}} \left| \sqrt{I_{real}(\mathbf{u})} - \sqrt{I_{ideal}(\mathbf{u})} \right|}{\sum_{\mathbf{u}} \sqrt{I_{ideal}(\mathbf{u})}}, \quad (1)$$

in which $I_{real}(\mathbf{u})$ is the simulated intensity at spatial frequency $\mathbf{u}$ including the effects of damage to the scattering target, $I_{ideal}(\mathbf{u})$ is the corresponding intensity distribution in the absence of damage, and the summations over $\mathbf{u}$ include all discrete samples included in the three-dimensional set of diffraction data.

The measure of data quality may be further refined, but its principal role is to provide a quantitative estimate of how closely the crystallographic model matches the experimental data. The requirement that $R \leq 0.15$ has been suggested[1] as a 'rule of thumb' by which diffraction data obtained from femtosecond XFEL experiments on isolated molecules should be regarded as possessing sufficient information to obtain a molecular structure with a spatial resolution determined by the maximum measured scattering angle. The $R$-factor has guided all computational studies regarding the pulse requirements for X-ray diffraction imaging of single biological molecules[1,4].

Implicit in the $R$-factor are the assumptions that the incident illumination and the diffracted wave possess full spatial and temporal coherence, independent of the nature of the matter-radiation interaction; these are implicit components of the crystallographic model against which the data are assessed. The validity of these assumptions is not supported, however, by a detailed consideration of the interaction physics that describes an encounter between a molecule and an intense XFEL pulse. This critical assessment pertains to the proposed experiments to determine molecular structures, even if the illumination exhibits full spatial and temporal coherence or if the scattering interaction takes place over so short a time that the nuclear framework is completely unaffected. A key component of the formulation in the present letter is the observation that the time-dependent evolution of the electron density imparts on the scattered wave the statistical characteristics of a partially-coherent wavefield, which we incorporate explicitly in our analysis.

Two critical assumptions are made in our approach. The first is that the collection of diffraction data originates from a matter-radiation interaction that is short enough in duration that we may assume that the nuclei are fixed in space throughout the encounter. Detailed simulations[1,5] put an upper limit on this period of 5 fs, which we assume is to be achieved either by pulse shaping or data gating. The second assumption is that a three-dimensional diffraction set has been assembled in the manner described, for example, of Fung *et al.*[3]. Each distinct molecular orientation must be associated with a sufficient number of two-dimensional molecular projections that that the stochastic distribution of molecular electronic vacancies created by photoionization and Auger emission may be represented by on-site statistical atomic averages. This is also an implicit assumption of the electrodynamical simulations performed by Hau-Riege *et al.*[5], whose general approach we have adapted to our purpose.

The time-dependent electron density is represented in the form

$$\rho(\mathbf{r},t) = \sum_{Z_\gamma} a_{Z_\gamma}(t) \sum_{m_Z} \rho_{Z_\gamma}\left(\mathbf{r} - \mathbf{R}^Z_{m_Z}\right). \qquad (2)$$

Since we are concerned with reproducing detailed molecular scattering properties rather than dynamical averages, we depart from the earlier treatment[8] by utilizing a model atomic basis of spherically-average orbital electron densities, $\rho_{Z_\gamma}\left(\mathbf{r} - \mathbf{R}^Z_{m_Z}\right)$, located at nuclear positions $\mathbf{R}^Z_{m_Z}$, where $Z$ labels the atomic species and $\gamma$ labels the orbital shell. The coefficients $a_{Z_\gamma}(t)$ define the time-variation in the occupancy of $\rho_{Z_\gamma}\left(\mathbf{r} - \mathbf{R}^Z_{m_Z}\right)$, averaged over all equivalent sites. This representation is augmented by continuum approximation for the density of electrons trapped by the residual ionic charge of the molecule, which is treated as a separate species.

The coherence properties of the scattered wavefield are incorporated in a matrix, $\mathbf{A}$, whose elements are defined by the statistical averages

$$A_{Z_\gamma, Z'_{\gamma'}} = \left\langle I(t) a_{Z_\lambda}(t) a_{Z'_{\lambda'}}(t) \right\rangle, \qquad (3)$$

where $I(t)$ represents the time-dependent intensity of the incident pulse. The integrated intensity, $I(\mathbf{q})$, corresponding to momentum transfer, $\mathbf{q}$, may be written in the form

$$I(\mathbf{q}) = \sum_{Z}\sum_{Z'} T_Z(\mathbf{q}) A_{Z,Z'}(q) T_{Z'}^*(\mathbf{q}), \quad (4)$$

where

$$T_Z(\mathbf{q}) = \sum_{m_Z} \exp\left(-i\mathbf{q} \bullet \mathbf{R}_{m_Z}^{Z}\right), \quad (5)$$

$q = |\mathbf{q}|$, $A_{Z,Z'}(q) = \sum_{\gamma,\gamma'} A_{Z_\gamma, Z'_{\gamma'}} f_{Z_\gamma}(q) f_{Z'_{\gamma'}}(q)$ and $f_{Z_\gamma}(q)$ is the spherically-symmetric elastic X-ray scattering factor for $\rho_{Z_\gamma}(\mathbf{r})$. In this model, all of the structural information is contained within the vector $T_Z(\mathbf{q})$, and all electrodynamical information is contained within $A_{Z,Z'}(q)$; one may determine one of these quantities from measurements of $I(\mathbf{q})$ if the other is known to sufficient accuracy. Our attention here is restricted to the determination of molecular structures, but we note that $A_{Z,Z'}(q)$ is a smooth, continuous function of $q$, so that electronic damage may also be characterized from measurements of $I(\mathbf{q})$ in systems for which $T_Z(\mathbf{q})$ is known.

Under the interaction conditions of interest, $A_{Z_\gamma, Z'_{\gamma'}}$ is poorly approximated by $A_{Z_\gamma, Z'_{\gamma'}} \simeq c_{Z_\gamma} c_{Z'_{\gamma'}}$, for fixed constants $c_{Z_\gamma}$ and $c_{Z'_{\gamma'}}$. As a consequence, $I(\mathbf{q})$ possesses the statistical characteristics of a partially-coherent diffraction pattern and the *R*-factor provides an inappropriate measure of its information content.

It has recently been demonstrated[8] that explicit incorporation of models of partial coherence into the solution of inverse problems may dramatically improve the quality of reconstructions using iterative, propagation-based techniques. The extension of this approach to the present case leads us to write $I(q)$ as a modal expansion of the form[9]

$$I(\mathbf{q}) = \sum_{k=1}^{\infty} \eta_k \psi_k(\mathbf{q}) \psi_k^*(\mathbf{q}) \qquad (6)$$

where the real, non-negative parameters, $\eta_k$, and the corresponding modes, $\psi_k(\mathbf{q})$, are solutions of an integral equation defined by the far-field mutual optical intensity. These modes are the Fourier transforms of a complete, orthonormal set of charge distributions that represent the time-averaged molecular charge density. The coherent diffraction limit is recovered in the limit $\eta_1 > 0$ and $\eta_k = 0$ for $k \geq 2$, in which case $\psi_1(\mathbf{q})$ corresponds to the usual crystallographic form-factor. A scheme to calculate the modal expansion under typical experimental conditions within this electrodynamical model appears in the Methods section.

An efficient structure determination algorithm that incorporates explicitly the effects of electronic damage and the partially-coherent nature of the scattering process may be formulated by adopting a change of basis and performing the generalized similarity transformation

$$I(q) = T(q)B(q)T^*(q) \qquad (7)$$

The structure factor, $T(q) = \sum_Z Z T_Z(q)$, directly encodes the structural parameters, $\mathbf{R}_{m_Z}^Z$, in the nuclear density, rather than the electron density. The matrix $B(q)$ must be approximated in such a manner that it preserves the statistical character of the electrodynamical information within $A_{Z,Z'}(q)$ and the chemical composition of the target molecule; this construction is described in the Methods section. The matrix, $B(q)$, and the imposition of a molecular support constraint define a conventional single-mode inverse problem for the molecular structure, essentially independent of the extent of damage.

Three-dimensional diffraction data for bacteriorhodopsin were generated using the electrodynamical model[5] described in Methods, assuming that $10^{12}$ 10keV photons were incident on the target in each 5 fs pulse. Compared with a similar calculation in the absence of any electronic damage, this resulted in an R-factor of m.n. Figure 1 shows samples of the undamaged diffraction pattern (Fig 1A) and damaged pattern (Fig 1B). This data displays the reduction in the high-$q$ scattering resulting from depletion of the core electrons through the pulse. Figure 1C shows the ratio of these two patterns and illustrates that the impact of the electronic damage imposes an uneven and structure dependent modification to the diffraction data indicating that a simple re-scaling of the pattern[10], though appealing, is not adequate in practice.

The Methods section describes the reconstruction algorithm in detail. A projection through the reconstruction is shown in Figure 2, where it can be seen that the atomic number and location of each atomic species is recovered essentially perfectly. Indeed, each atomic position may be located beyond the formal resolution of the diffraction pattern using a centroiding process on the fringes around each nucleus; if the nucleus is centrally located in each pixel then these fringes are absent and their relative amplitudes encode the location of the nucleus is encoded in the form of the fringes. The concept of super-resolving crystallographic data has recently emerged in a rather different but related context[11]. There is no need to construct a representation of $\rho(r,0)$, or to determine $R^Z_{m_Z}$ from it by crystallographic model-building. This appears to be the first reconstruction of its type; the molecular structure of a complex biomolecule has been recovered directly under interaction conditions in which the electron density is so comprehensively damaged that the usual working rules of protein crystallography are no longer valid. To underline this point, Figure 2C displays a detail from a slice through the reconstruction, demonstrating that the atomic species information is automatically recovered using this algorithm.

The application of the X-ray free electron laser to the determination of molecular structure is an extremely exciting endeavour. But it is also one that offers very considerable scientific and technical challenges. A very fundamental challenge is the need to understand the role of the interaction of the molecule with the incident field. The present letter demonstrates that recovery of molecular structure in the presence of damage is most readily achieved by adopting a model that reflects the detailed interaction physics, rather than the usual crystallographic assumption of full coherence. Our approach is predicated on the need for the damage mechanisms to be well characterized, either by kinetic modeling, or by inversion of diffraction data obtained from scattering targets of known structure. As such, we submit an important obstacle in the realization of single molecule structural biology using XFEL sources has been removed.

*This research was supported by the Australian Research Council through its Centres of Excellence and Federation Fellowships programs.*

*HQ performed the majority of the theoretical analysis and computational implementation. Both authors contributed to the conceptual formulation of the research and to the writing of the letter.*

*The authors do not have any competing interests.*

**METHODS**

The three-dimensional intensity distribution of X-ray photons scattered from a molecule of bacteriorhodopsin, *I(q)*, was simulated using the methods previously published by Hau-Riege et al.[5]

The modal distributions defined by Eq. (6) are determined in the following manner. A non-orthogonal basis of orbital densities with which to expand the modes is

$$\Phi_{Z\gamma}(r) = \sum_{m_Z} \rho_{Z\gamma}\left(r - R_{m_Z}^Z\right), \quad (8)$$

so that $\psi_k(\mathbf{r}) = \sum_{Z\lambda} c_k^{Z\gamma} \Phi_{Z\gamma}(\mathbf{r})$, and $c_k^{Z\gamma}$ is an associated expansion coefficient which, in the present context, must be a real quantity. These coefficients and the corresponding real and non-negative eigenvalues, $\eta_k$, that appear in Eq. (6) are determined as the solutions of a generalized matrix eigenvalue equation of the form $\mathbf{JC} = \mathbf{\eta SC}$. The diagonal matrix, $\mathbf{\eta}$, contains the eigenvalues, the columns of C contain the eigenvectors and the matrix elements of **J** and **S** are defined by

$$J_{Z_1\gamma_1, Z_2\gamma_2} = N_{Z_1} N_{Z_2} \sum_{\gamma_3\gamma_4} \Psi_{\gamma_1,\gamma_3}^{Z_1} A_{Z_1\gamma_3, Z_2\gamma_4} \Psi_{\gamma_4,\gamma_2}^{Z_2}, \quad (9)$$

$$S_{Z_1\gamma_1, Z_2\gamma_2} = N_{Z_1} \Psi_{\gamma_1\gamma_2}^{Z_1} \delta_{Z_1 Z_2}, \quad (10)$$

where $N_Z$ is the number of atoms of nuclear charge $Z$, $\delta_{ZZ'}$ is the Kroneker delta, and

$$\Psi_{\gamma_1\gamma_2}^Z = \int \rho_{Z\gamma_1}(\mathbf{r}) \rho_{Z\gamma_2}(\mathbf{r}) d\mathbf{r}. \quad (11)$$

For consistency with the diffraction simulations, the orbital density integrals were evaluated using a spherical-atom model derived from Slater-type orbitals[12] within the tight-binding approximation. We note, however, that the formalism is readily extended to more detailed numerical treatments of the electron density simply by interpreting γ to be a label identifying *N*-electron electronic state functions including the effects of electronic relaxation, correlation and relativistic corrections.

In order to reconstruct molecular structure from diffraction data including electronic damage, we introduce a function, $B(\mathbf{q})$, with the property

$$\sum_{Z}\sum_{Z'} T_Z(\mathbf{q}) A_{Z,Z'}(\mathbf{q}) T_{Z'}^*(\mathbf{q}) = B(\mathbf{q}) T(\mathbf{q}) T^*(\mathbf{q}). \qquad (12)$$

This relationship is used to facilitate a change of basis. The left-hand side of the above equation, which models the partially-coherent diffraction data, $I(\mathbf{q})$, is modified to form $T(\mathbf{q})T^*(\mathbf{q}) \simeq I(\mathbf{q})/B(\mathbf{q})$, which contains only information about the positions of the nuclei within the molecule; all electronic information is removed from $I(\mathbf{q})$ by $B(\mathbf{q})$.

We require that the diagonalizing transformation, $B(\mathbf{q})$, introduces no significant bias towards any particular molecular structure. It is estimated using its definition,

$$B(\mathbf{q}) = \frac{\sum_{ZZ'} T_Z(\mathbf{q}) A_{ZZ'}(q) T_{Z'}^*(\mathbf{q})}{\sum_{ZZ'} T_Z(\mathbf{q}) T_{Z'}^*(\mathbf{q})},$$

in which it should be noted that any rapid variation in the unknown products of the form $T_Z(\mathbf{q})T_{Z'}^*(\mathbf{q})$ in the above expression will cancel, to a good approximation, because $A_{ZZ'}(q)$ is a smooth, slowly-varying function of *q*. Since we are particularly interested in biomolecules, a simple model in which the discrete nuclear positions are replaced by continuous spherical uniform charge distributions of finite radius, *R*, provides an appropriate approximation scheme. We find that, within this model,

$$T_Z(\mathbf{q})T_{Z'}^*(\mathbf{q}) = \frac{9 N_Z N_{Z'}}{(qR)^6} \left\{ \begin{array}{c} [2qR\cos qR - (2-q^2R^2)\sin qR]^2 \\ + [2qR\sin qR - (2-q^2R^2)\cos qR - 2]^2 \end{array} \right\}, \qquad (13)$$

where $N_Z$ is the number of atoms of charge Z, and $q = |\mathbf{q}|$. For $q \to 0$, this function has the limit $N_Z N_{Z'}$, and vanishes for $q \to \infty$, because no two atoms of different charges may share the same spatial position. In the diagonal case, $Z = Z'$, we require that the function should possess the limit $N_Z$ for $q \to \infty$, and so a correction of the form $N_Z \left[1 - \exp(-\mu q^2)\right]$ is added to this approximation. The selection of the value of $\mu$ proves to be not very critical, and we use $\mu \simeq R$. The above approximation for $B(\mathbf{q})$ is used to define a modified inverse problem whose solution is $T(\mathbf{q})$, and which satisfies a finite support constraint and Nyquist sampling requirements as the determination of molecular electron density from diffraction data., $I(\mathbf{q})$, including the effects of electronic damage.

The solution of the resulting inverse problem is essentially conventional, and employs the iterative hybrid input-output and error-reduction algorithms described by Fienup, supplemented by the judicious use of 'charge-flipping' in the early stages of the structure recovery to establish the effective exterior support surface of the molecule. The iterations are initialized by a uniform charge distribution with a radius of 30 Å, and typically converge within 1000 iterations.

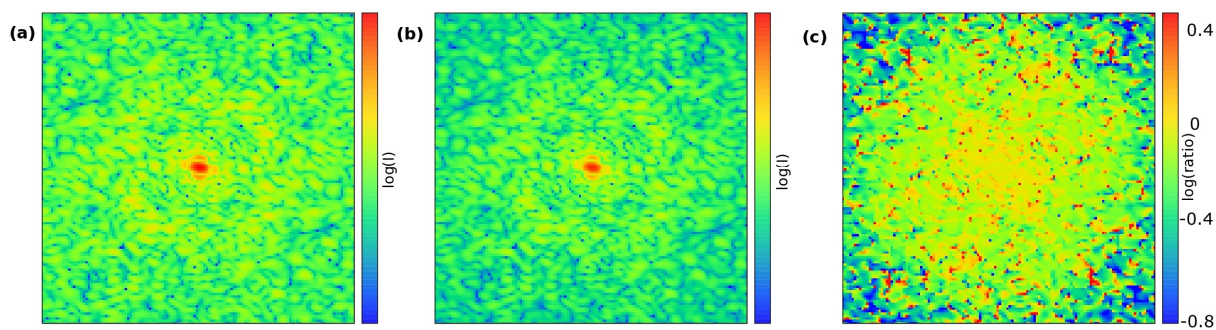

FIG 1. A full three-dimensional diffraction pattern from the bacteriorhodopsin molecule is used in this work, simulating the form of data that would be recovered after a successful alignment of the data from XFEL interactions with many randomly-orientated molecules. (a) The diffraction pattern from the molecule with the damage mechanisms turned off. (b) The diffraction pattern obtained when the molecule is illuminated by a 5 fs pulse containing $10^{12}$ photons and based on the damage mechanisms described in the text. (c) The logarithm of the ratio of the damaged to undamaged pattern, normalized so that the intensities at $|q|=0$ are of equal value. The reduction in the scattered intensity at high diffraction angles is apparent. Note that the ratio is highly non-uniform, so that a simple re-scaling is not able to recover the correct pattern. Adjacent red and blue pixels indicate regions where the ratio diverges indicating that a zero in the diffraction pattern has moved due to the effects of damage; there are many such regions throughout the data. The width of the array in these calculations corresponds to a spatial resolution of 1.04 Å.

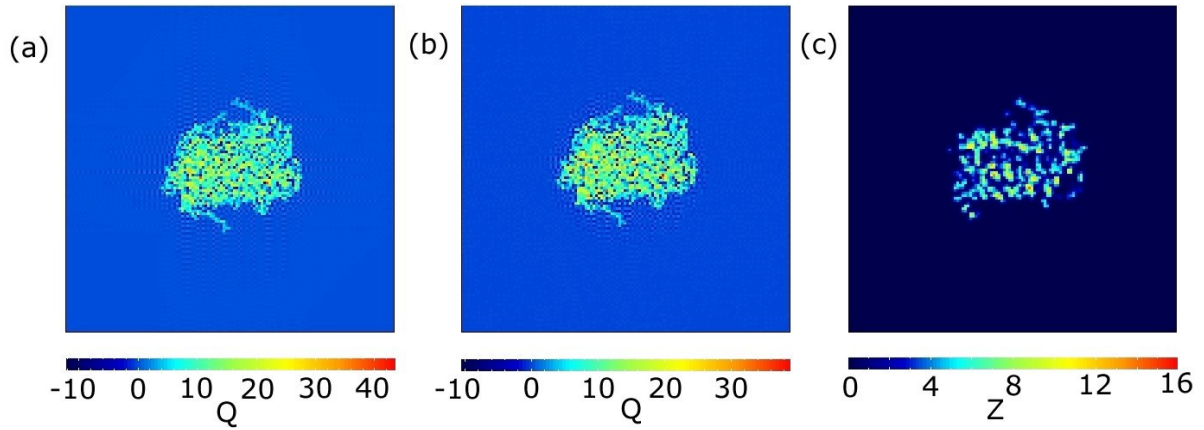

FIG 2. (a) Two-dimensional projection, Q, of the nuclear charge (in atomic units) through bacteriorhodopsin. It is evaluated as the Fourier transform of $T(\mathbf{q})$, which is constructed directly from its definition and that of $T_Z(\mathbf{q})$, Eq. (5), using the nuclear positions of bacteriorhodopsin obtained from the Protein Data Bank. The visible structure outside the molecular envelope carries super-resolution information that is available for recovery from high-resolution XFEL diffraction data that reflects the fact that the nuclear positions do not generally sit at the central pixel position. For comparison purposes, each nuclear position is assigned the value of the nuclear charge at that point. (b) The reconstruction of the projected charge, Q, from simulated diffraction data corresponding to the interaction of bacteriorhodopsin with a 5 fs XFEL pulse of $10^{12}$ 5 keV photons. The weak attenuation of the super-resolution fringes by the spherical support function is visible; the reconstruction corresponds to the input structure in all significant structural details and confirms that the approach described here recovers the full structural information. (c) Detail of a slice through the full three-dimensional reconstruction illustrating the recovery of the details of the molecular structure. In this case, the nuclear charge in the slice is plotted. One can directly associate a particular atomic species with a given site in the molecule by simply reading off the scale. To aid visualization, the fringes were eliminated from this plot by convolving the reconstruction with a three-dimensional Gaussian function.